\pdfoutput=1

\documentclass[11pt]{article}

\usepackage[final]{acl}

\usepackage{times}
\usepackage{latexsym}
\usepackage{pifont}

\usepackage[T1]{fontenc}

\usepackage[utf8]{inputenc}

\usepackage{microtype}

\usepackage{inconsolata}

\usepackage{graphicx}

\usepackage{algorithm}
\usepackage{algpseudocode} 

\usepackage[hypcap=false]{caption}
\usepackage{cuted}
\usepackage{amsmath, amssymb, bm, geometry}
\usepackage{caption}
\usepackage{subcaption}
\usepackage{listings}
\usepackage{enumitem}
\usepackage{booktabs}
\usepackage{multirow}
\usepackage[most]{tcolorbox}
\title{MobileCity: An Efficient Framework for\par
Large-Scale Urban Behavior Simulation}

\author{
Xiaotong Ye\textsuperscript{1}\thanks{Equal contribution.} \quad Nicolas Bougie\textsuperscript{1}\footnotemark[1] \quad Toshihiko Yamasaki\textsuperscript{2} \quad Narimasa Watanabe\textsuperscript{1} \\
\textsuperscript{1}Woven by Toyota \\
\textsuperscript{2}The University of Tokyo \\
\texttt{\{tony.yip, nicolas.bougie, narimasa.watanabe\}@woven.toyota} \\
\texttt{yamasaki@cvm.t.u-tokyo.ac.jp}
}

\begin{document}
\maketitle
\begin{abstract}
Generative agents offer promising capabilities for simulating realistic urban behaviors. However, existing methods often rely on static profiles, oversimplified behavioral logic, and synchronous inference pipelines that hinder scalability. We present \textbf{MobileCity}, a lightweight generative-agent framework for city-scale simulation powered by cognitively-grounded generative agents. Each agent acts based on its needs, habits, and obligations, evolving over time. Agents are initialized from survey-based demographic data and navigate a realistic multimodal transportation network spanning multiple types of vehicles. To achieve scalability, we introduce asynchronous batched LLM inference during action selection and a low-token communication mechanism. Experiments with 4,000 agents demonstrate that MobileCity generates more human-like urban dynamics than baselines while maintaining high computational efficiency. Our code is publicly available at \href{https://github.com/Tony-Yip/MobileCity}{https://github.com/Tony-Yip/MobileCity}.

\end{abstract}

\section{Introduction}
Generative agents~\cite{uist/ParkOCMLB23}, powered by Large Language Models (LLMs)~\cite{emnlp/MadaanZ0YN22}, have emerged as a transformative paradigm for simulating human-like behaviors across domains including recommender systems~\cite{sigir/0003CSWC24}, peer review~\cite{bougie2024generative}, medical Q\&A~\cite{corr/abs-2405-02957}, and game simulation~\cite{corr/abs-2309-11672,corr/abs-2404-02039}. Urban simulation models behaviors and transportation within a city, enabling evaluation of policies, transportation changes, and infrastructure planning. It supports forecasting market demand, analyzing traffic and safety impacts, and assessing public health and community well‑being.

Despite recent progress~\cite{uist/ParkOCMLB23,emnlp/WangCC23}, existing generative-agent frameworks exhibit notable limitations for large-scale urban mobility simulation. Most systems do not explicitly model human needs, temporal routines, or obligation-driven behaviors, leading to repetitive or unrealistic activity patterns~\cite{feng2024citybench,samuel2024personagym,bougie-watanabe-2025-citysim}. Besides, prior work typically assumes a single or overly simplified transportation system and fail to incorporate environmental factors such as weather or temperature, limiting the realism of mobility decisions. Finally, synchronous LLM calls and multi-turn dialogues incur substantial token and computation costs, making them prohibitively expensive to run at scale~\cite{kaiya2023lyfe}.

In light of this, we introduce \textbf{MobileCity}, a scalable generative-agent simulator built on a tile-based city representation. Agents are initialized with survey-based demographic profiles and evolve according to dynamic internal states through three modules: a \textit{needs}, a \textit{habits}, and an \textit{obligations} module governing compulsory tasks. MobileCity further incorporates a multi-modal transportation system with three mobility options and integrates environmental factors such as weather, temperature, and venue availability, enabling context-aware decisions. Finally, in order to ensure scalability, we employ asynchronous batched LLM calls for action selection, streamline communication by exchanging memory indices instead of generating dialogues, and record only event-level state changes in an \texttt{OpenSearch} backend. Experiments with 4{,}000 agents demonstrate that MobileCity achieves higher behavioral realism and significantly better simulation efficiency compared to existing baselines. Beyond improving fidelity, we also showcase practical applications in mobility pattern forecasting and demographic analytics, illustrating MobileCity's utility for urban planning and computational social science.

Our main contributions are:
\begin{itemize}
\item \textbf{Cognitively-grounded, survey-conditioned urban agents.}
We propose MobileCity, where each agent’s behavior is jointly driven by needs, habits, and obligations, and initialized from survey-based demographic and behavioral profiles, enabling diverse and temporally realistic daily routines. 

\item \textbf{Multi-modal mobility and context-aware decision making.}
We incorporate a realistic transportation system with multiple modes and integrate environmental factors such as weather, temperature, and venue availability to support context-aware mobility and activity choices.

\item \textbf{A scalable, low-token simulation pipeline for thousands of agents.}
We achieve efficient city-scale simulation via (i) constrained, multiple-choice action selection to reduce token usage, (ii) lightweight communication by exchanging memory indices instead of generating full dialogues, and (iii) asynchronous batched LLM inference with event-level logging for scalable execution.
\end{itemize}

\section{Related Work}
Recent studies on generative agents have demonstrated significant progress in simulating human behavior. \citet{uist/ParkOCMLB23} introduce the first framework in which agents maintain memories and engage in social interactions. Building upon this foundation, \citet{emnlp/WangCC23} incorporate basic needs to make daily activities more realistic, while \citet{iclr/ChenSZ0YCYLHQQC24} design customizable environments that support emergent collaborative behaviors. To broaden applicability, \citet{corr/abs-2309-07870} present an open-source system for autonomous language agents, and \citet{iclr/HongZCZCWZWYLZR24} demonstrate how agents can collaborate in complex software engineering workflows.

As research moves toward larger-scale simulations, computational efficiency becomes a central concern. \citet{corr/abs-2411-10109} scale simulations to 1,000 agents through a hierarchical decision-making architecture, although the proposed architecture still incurs prohibitive inference costs. \citet{corr/abs-2402-02053} reduce unnecessary LLM calls by learning simplified policies, yet real-time simulations remain constrained by the latency of LLM responses, especially when generating multi-turn dialogues.

Despite these advancements, existing systems typically overlook several factors essential for realistic urban mobility: diverse transportation modes, environmental conditions such as weather or temperature, and long-term behavioral traits influenced by needs, habits, and obligations. Moreover, prior work \cite{bougie-watanabe-2025-citysim} usually relies on token-intensive content generation. As a result, generating human-like behaviors with low inference cost and high scalability remains an open challenge.

\section{Agent Modules}
\subsection{Agent Profile}
We derive personas from questionnaire surveys completed by human participants, enabling the simulation to capture diverse demographic and psychological characteristics. Each agent is initialized with the following attributes:
\begin{itemize}
    \item \textbf{Demographic Information} includes gender, age, job category, eduction level, financial status, family status like marriage.
    \item \textbf{Human Parameters}~\cite{barrick1991big} describe long-term behavioral tendencies. They include the Big Five personality traits and behavioral traits.
    \item \textbf{Hobbies} are initialized from SNS data, like X Posts, and dynamically updated based on agents’ activity records during simulation.
\end{itemize}

\subsection{Individual Action Module}
\label{sec:action_modules}
Human decisions arise from three psychological mechanisms that drive human action~\cite{wood2022habits}: \textit{needs} ("I want to do"), \textit{habits} ("I do it as usual"), and \textit{obligations} ("I have to do"). We formalize them into three separate modules.

\subsubsection{Needs-driven Action}
Agents have spontaneous tendencies to maintain physiological or social equilibrium, consistent with theories of homeostasis~\cite{cannon1932wisdom}, and interpersonal balance~\cite{festinger1957cognitive,heider1958psychology}. Namely, when an agent's internal state deviates from its optimal level, it seeks to restore or enhance that state. We introduce eight agent needs, following ~\citet{maslow1943theory}’s hierarchy of needs in Table~\ref{table:Maslow Needs}.

\begin{table}[h]
\label{table:agent_needs}
\resizebox{1.0\columnwidth}{!}{
\begin{tabular}{lr}
\toprule
\textbf{Maslow's Hierarchy} & \textbf{Agent Needs} \\
\midrule
Physiological & Fullness, Energy \\
Safety & Health, Financial Security \\
Love/Belonging & Pleasure, Social Connection \\
Esteem & Status Recognition \\
Self-Actualization & Self-Growth \\
\bottomrule
\end{tabular}
}
\caption{
Maslow ’s hierarchy of needs}
\label{table:Maslow Needs}
\end{table}

Each agent maintains a vector of need levels $C_N \in [0,1]^8$, which decays over time following $C_N(t + \Delta t) = \mathrm{clip}\big(C_N(t) - \Delta t \cdot D_N,\; 0,\; 1\big)$ , where $D_N$ represents the individual decay rate vector. In contrast to prior work \cite{bougie-watanabe-2025-citysim, yan2024opencity}, decay rates are heterogeneous across personas and need types. Lower-level physiological needs decay faster, while higher-level needs are more stable. For example, residents living alone experience quicker decline in \textit{Social Connection} due to increased susceptibility to loneliness. In addition, we maintain an importance vector $I_N$, which encodes how strongly an agent prioritizes each need. For instance, agents with lower income place higher importance on financial security. 

During action selection, the needs-driven score of an action at time \(t\) is defined as $N(t) = N_{\mathrm{hp}} \, N_{\mathrm{imp}}$. \(N_{\mathrm{hp}}\) represents the weighted cosine similarity score between the agent's human parameters \(x_{\mathrm{hp}}\) and the action's feature vector \(x_{\mathrm{act}}\). \(N_{\mathrm{imp}}\) measures the importance-weighted fulfillment of unsatisfied needs defined by $1-C_N$ and $I_N$. 

\subsubsection{Habit-driven Action}

Habit-driven actions are triggered by temporal or spatial regularities reinforced through repeated actions. To reproduce such patterns, we define a habitual action preference function. 

Suppose that the agent performed an action in the past, with the midpoint time of $t_m$, the amplitude, defined by action feedback, is $A_H$. We aim to determine the habit strength at the current time $t$. To model the daily cycle on a continuous circular domain, we normalize the minute-based time difference onto the interval $[-\pi, \pi]$ using $\Delta\theta(t)
= \tfrac{2\pi}{1440}\,\big((t - t_m) \bmod 1440\big)$, where $\Delta\theta(t)$ is the normalized angular distance. The habit intensity as a function of current time is modeled as a Gaussian distribution on the circle: $H(t) = R(t)A_H \, \exp\!\big(-\,k_H\,\Delta\theta(t)^{2}\big)$, where $k_H$ controls the sharpness of the temporal peak, which is defined by the angular half-width of action execution time $a_H$, and $A_H$ is defined by $k_H$ to maintain a constant area. $R(t)$ represents the forgetting strength in the Ebbinghaus~\cite{rubin1996one} forgetting curve model. As time passes, the habit strength will gradually decrease by $R(t) = \exp\!\big(-\,r_H\,(t-t_m)\big)$. Habits are removed entirely once their strength falls below a minimal threshold.

\subsubsection{Obligation-driven Action}
Obligation-driven action refers to behaviors selected not from internal needs or habits but from externally imposed duties~\cite{gershuny2003changing}. In our framework, these mandatory tasks are encoded as core time slots in each agent’s calendar, derived from our questionnaire survey. They reflect factors such as sleep schedules, family structure (e.g., cohabitation, marital status, children’s ages), and historical activity logs.

During action selection, candidate needs-driven and habit-driven actions are first filtered by an availability mask determined by the next mandatory task. An action is admissible only if: a) it is semantically appropriate for the current time (e.g., “eat breakfast’’ is invalid at night), b) its venue is open during the intended period, and c) the agent can complete it, including travel time, and still arrive at the upcoming mandatory task on schedule.

\subsection{Mobility Selection Module}
\label{sec:mobility_selection}
When the locations of an agent’s consecutive actions differ, the agent must choose an appropriate mode of transportation. We implement a transportation system within the virtual town, comprising three transportation modes: walking, PMV (personal mobility vehicle), and bus. During action selection, the LLM is instructed to select an action from a list of multi-mechanism-driven actions and the most appropriate transportation mode, conditioning on the agent's persona and environmental information including weather, temperature, and spatial context.

\section{Towards Scalable Simulation}
\label{sec:scalable_simulation}
One of the primary goals of our system is to enable efficient simulation of large-scale agent populations. To this end, we introduce three strategies to improve efficiency.

\subsection{Reducing Token Consumption}
We first reduce token usage in the individual action module by shifting the LLM’s role from free-text generation to discrete selection. Specifically, the \textbf{action selector} precomputes a list of feasible candidate actions ${Act_{needs}, Act_{habit}, Act_{obl}}$, with the mechanisms described in Section~\ref{sec:action_modules}, and mobility options $walking, PMV, bus$ in Section~\ref{sec:mobility_selection}. The LLM is then prompted with a multiple-choice question containing these candidates, and its output is restricted to the index of the chosen option. An example is illustrated in Appendix~\ref{appendix_action_selector}.

Instead of generating full dialogues, agents exchange information through a lightweight memory-transfer mechanism. The LLM is prompted to select which memory entries are shared between agents and to update their mutual relationship scores. Formally, when agents $i$ and $j$ meet, the LLM outputs only: $(\Delta \mathcal{M}_i, \Delta \mathcal{M}_j, \Delta R_{ij}) = \mathrm{LLM}_{\text{comm}}\big(\mathcal{M}_i, \mathcal{M}_j, \text{context}_{ij}\big)$, where $\Delta \mathcal{M}$ represents the exchanged memory indices, and $\Delta R_{ij}$ updates the social affinity between agents.

\subsection{Asynchronous Mechanism}
A central component of our scalability strategy is asynchrony. Our city-scale agent simulator operates under an asynchronous scheduling mechanism. At the beginning of each simulated day, a list of all agents $\mathcal{A} = \{a_1, a_2, \dots, a_N\}$ is initialized, and the system maintains a set of independent local clocks $\mathcal{I} = \{\theta_1, \theta_2, \dots, \theta_N\}$. This design allows each agent to progress through its own timeline, rather than synchronizing with a global simulation step. The pseudo-code is shown below.

\begin{center}
\small
\begin{tabular}{p{0.95\linewidth}}
\toprule
\textbf{Asynchronous Action Batch Scheduling} \\
1. Initialize agents $A$, clocks $\theta$, batch $\mathcal{B}$, threshold $B$.\\
2. For each $a_i\in A$: \\
\quad (a) If mandatory task due $\to$ execute and advance $\theta_i$.\\
\quad (b) Else compute candidates from needs and habits, append to $\mathcal{B}$.\\
3. If $|\mathcal{B}|=B$ or all awaiting $\to$ dispatch batch.\\
4. Update $(\theta_i,C_N)$ for returned agents.\\
5. Remove agents with $\theta_i\ge24{:}00$. Repeat until $A=\emptyset$.\\
\bottomrule
\end{tabular}
\vspace{0.3cm}
\end{center}

The same mechanism is applied to agent-to-agent communication. Throughout the simulation, pairs of agents $(a_i, a_j)$ likely to converse are dynamically generated, or when agents proactively reaching out when their social need is high. Instead of invoking the language model for every pair immediately, the system collects communication tasks into a shared batch buffer. Once the batch reaches a predefined threshold, all pending conversations are processed asynchronously, exchanging memory indices and updating relationship scores in parallel:

\begin{center}
\small
\begin{tabular}{p{0.95\linewidth}}
\toprule
\textbf{Asynchronous Conversation Batch Scheduling} \\
1. Initialize conversation batch $\mathcal{B}_{\text{conv}}$, threshold $B_{\text{conv}}$.\\
2. Detect potential pairs $(a_i,a_j)$: \\
\quad (a) Face-to-face if both share venue and time overlap.\\
\quad (b) Virtual contact if agent $a_i$ has high social need.\\
3. Append $(a_i,a_j,\textsc{Memory}_i,\textsc{Memory}_j)$ to $\mathcal{B}_{\text{conv}}$.\\
4. If $|\mathcal{B}_{\text{conv}}|=B_{\text{conv}}$ $\to$ dispatch batch.\\
5. LLM returns exchanged memories and relationship updates 
$(\Delta\mathcal{M}_i,\Delta\mathcal{M}_j,\Delta R_{ij})$.\\
6. Update memories and relationship states.\\
\bottomrule
\end{tabular}
\vspace{0.3cm}
\end{center}

\subsection{Data Logging and Visualization}
In previous simulation systems~\cite{uist/ParkOCMLB23, emnlp/WangCC23}, the state and location of all agents at every time step were saved into local \texttt{JSON} files, which were then repeatedly accessed by the frontend for visualization. This I/O-intensive pipeline introduced significant latency and storage overhead. To address this issue, we decouple the simulation backend from the frontend and record only essential state changes. Specifically, each agent’s need satisfaction vector $C_N$ is logged only when an action is completed, and spatial coordinates are recorded only upon movement. All event-level logs are stored in an \texttt{OpenSearch}~\cite{opensearch-github} database instead of local files. After the simulation, missing agent states are linearly interpolated based on the temporal continuity of needs and locations, allowing the frontend to reconstruct smooth and continuous trajectories directly from \texttt{OpenSearch} queries.

\begin{figure*}[t]
\begin{center}
     \centering
     \begin{subfigure}[b]{\textwidth}
         \centering
         \includegraphics[width=15cm]{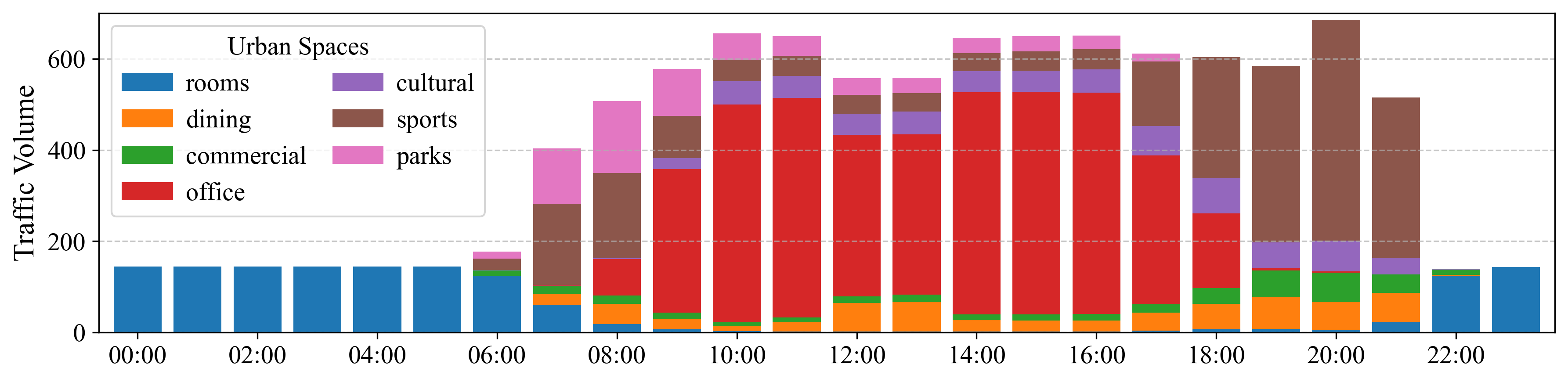}
         \label{fig:weekday_traffic}
     \end{subfigure}
     \begin{subfigure}[b]{\textwidth}
         \centering
         \includegraphics[width=15cm]{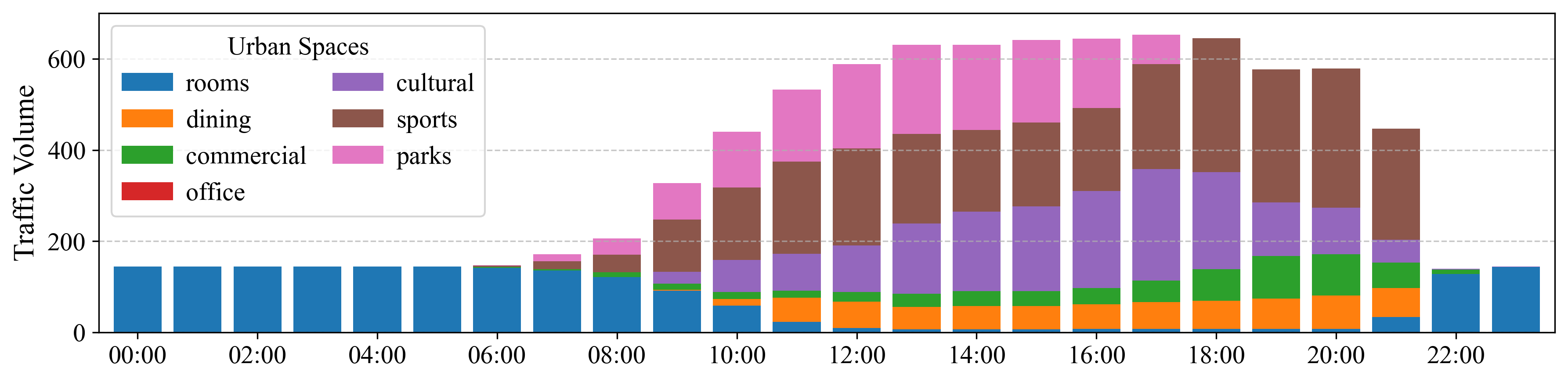}
         \label{fig:weekend_traffic}
     \end{subfigure}
     \caption{The crowd distribution across different urban venues, on weekdays (top) and weekends (bottom).}
     \label{fig:traffic}
\end{center}
\end{figure*}

\section{Experimental Results}
\label{sec:exp}
\subsection{Runtime Analysis}
Prior generative-agent systems suffer from severe runtime limitations due to heavy LLM invocation. Humanoid Agents~\cite{emnlp/WangCC23} requires 40 minutes to simulate only 3 agents. AgentSociety~\cite{gershuny2003changing} adopts cohort-based batching, yet inference for 1,000 agents partitioned into 8 groups still takes 
11 minutes for a single global decision cycle. 
These baselines highlight the computational bottleneck of LLM-driven multi-agent simulations and motivate the need for a more efficient execution framework. We accelerate end-to-end simulation using three mechanisms as explained in Section~\ref{sec:scalable_simulation}. 
To quantify their effects, we conduct an ablation analysis.

\begin{table}[h]
\centering

\begin{subtable}{0.9\linewidth}
\centering
\small
\begin{tabular}{lccc}
\toprule
Population & R & R+D & R+A+D \\
\midrule
40   & 194      & 115       & 39      \\
400  & 2,093    & 1,329     & 383     \\
4,000 & 22,432   & 15,234    & 3,734   \\
\bottomrule
\end{tabular}
\caption{Weekday}
\label{tab:runtime-weekday}
\end{subtable}

\begin{subtable}{0.9\linewidth}
\centering
\small
\begin{tabular}{lccc}
\toprule
Population & R & R+D & R+A+D \\
\midrule
40   & 246      & 154       & 52      \\
400  & 2,497    & 1,649     & 495     \\
4,000 & 29,656   & 20,731    & 4,850   \\
\bottomrule
\end{tabular}
\caption{Weekend}
\label{tab:runtime-weekend}
\end{subtable}
\caption{
Runtime (seconds) under different acceleration settings.
R = Reducing Token Consumption; 
R+D = adding Data Logging; 
R+A+D = full system including the Asynchronous Mechanism.
}
\label{table:runtime-sub}

\end{table}
We observe that weekday simulations consistently finish faster than weekend simulations. This is expected: employed agents spend a larger portion of weekday daytime in workplaces, resulting in fewer action selections and correspondingly fewer LLM calls for memory updates. In contrast, weekend schedules involve more frequent transitions across leisure venues, increasing the total number of model queries.

\subsection{Human Likeness}

A central question is how closely synthetic residents resemble real human behavior. 
To evaluate this, we present each agent’s daily actions to GPT-4o~\cite{king2023conversation} and ask it to judge whether the behavior appears human-like or machine-generated using a 5-point Likert scale. 
Higher scores indicate stronger alignment with natural human behavior. 
Table~\ref{table:human_like} reports the averaged scores across interactions, comparing our method with the baseline~\cite{uist/ParkOCMLB23}, AGA~\cite{corr/abs-2402-02053}, and HumanoidAgent~\cite{emnlp/WangCC23}. 
Our approach achieves the highest human-likeness score by a notable margin, demonstrating that the integration of needs, temporal habits, and obligation-driven decision-making produces behaviors that GPT-4o reliably interprets as human. 
A qualitative example of generated daily interactions is provided in Appendix~\ref{sec:action_example}.

\begin{table}[h]
\caption{Human-likeness score evaluated by GPT-4o.}
\label{table:human_like}
\begin{center}
\begin{tabular}{lc}
\toprule
Method & Activity \\
\midrule
Baseline & 3.11 ± 0.18 \\
AGA & 3.22 ± 0.28 \\
HumanoidAgent & 3.30 ± 0.31 \\
Ours & \textbf{4.09} ± \textbf{0.27} \\
\bottomrule
\end{tabular}
\end{center}
\end{table}

\begin{figure*}[t]
    \centering
    \includegraphics[width=15cm]{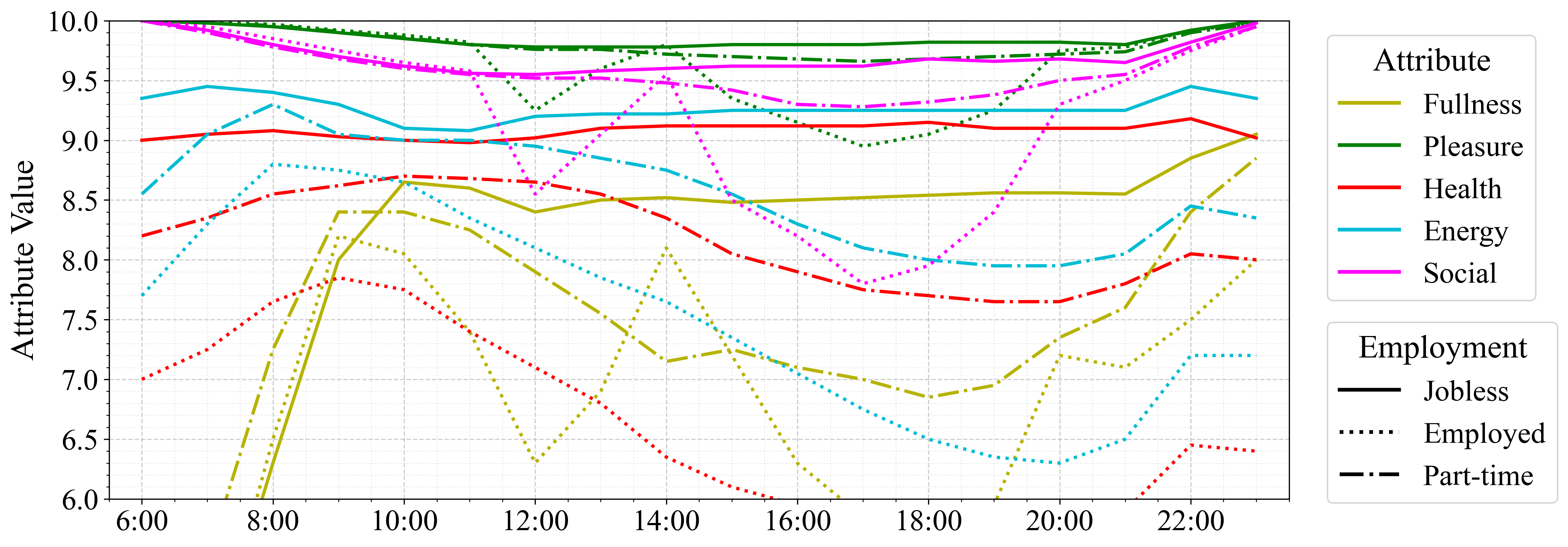}
    \caption{Residents with different employment status have different fluctuations in agent needs during the day.}
    \label{fig:basic_needs}
\end{figure*}

\subsection{Venue Heatmap}
Understanding how crowds occupy urban spaces over time is crucial for urban planning and resource allocation. 
Figure~\ref{fig:traffic} illustrates the temporal distribution of venue utilization generated by MobileCity. 
Between 22:00 and 06:00, most agents remain in residential rooms, reflecting natural nighttime resting patterns. 
During weekday mornings, employed agents concentrate in office areas, producing a pronounced surge in workplace occupancy. 
As work hours end, the population gradually shifts toward leisure-oriented venues such as sports centers, cultural spaces, and parks. 
In contrast, weekend patterns exhibit a more diverse distribution throughout the day, with consistent increases in visits to commercial, dining, and recreational locations. 
Overall, the simulated dynamics closely align with real-world urban mobility trends, where work schedules, leisure routines, and daily rhythms jointly shape venue occupancy.

\subsection{Macro-Level Action Distribution}
\begin{figure}[h]
    \centering
    \includegraphics[width=\columnwidth]{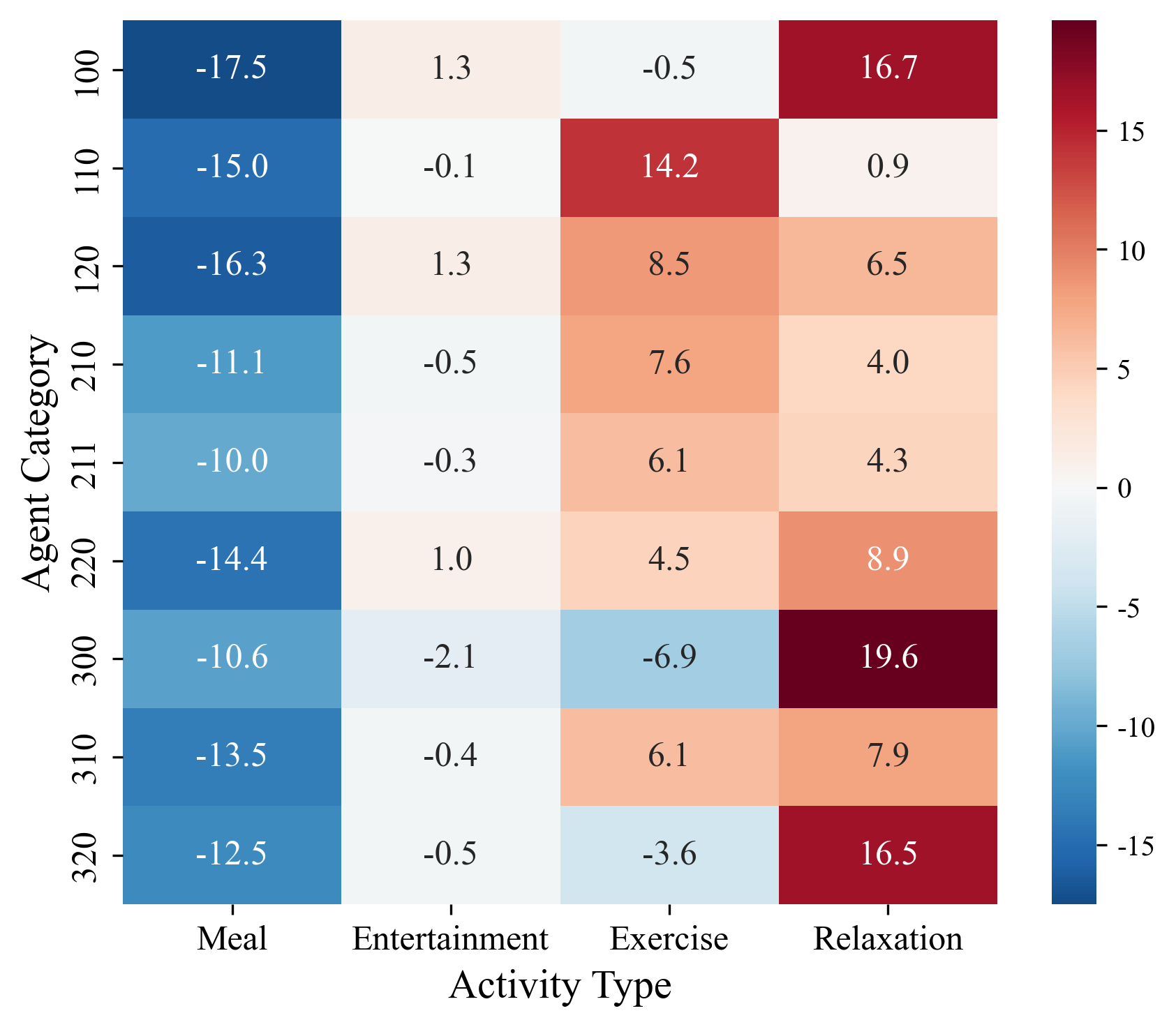}
    \caption{Percentage point differences in activity distribution between our method and real-world data across demographic categories.}     
    \label{figure:macro_results}
\end{figure}

While aligning individual agent behaviors with their human counterparts is crucial, it is also necessary that human proxies replicate real-world user behavior at a macro level. In each category, the three-digit code represents age, employment status, and income level. We compare the percentage distribution of activities between our method and real-world data. Figure \ref{figure:macro_results} presents this comparison as a heatmap of percentage point differences. Categories are encoded as three-digit IDs $XYZ$, where 
$X \in \{1,2,3\}$ denotes age group (1: 25--44, 2: 45--64, 3: 65--84), 
$Y \in \{0,1,2\}$ denotes employment status (0: unemployed, 1: employed, 2: part-time), 
and $Z \in \{0,1\}$ denotes income level (0: medium, 1: high). Unemployed agents have more time to perform those actions than employed agents, since employees have to work in the office on weekdays. We also noticed that employed adults show higher exercise engagement in our simulation, while older demographics exhibit shifted time allocation preferences. The observed differences provide valuable insights into demographic-specific behavioral tendencies that can inform future social studies while demonstrating our method's capability to replicate complex human behavioral patterns.

\subsection{Emotion Monitoring}
To analyze agents’ emotional states, we group agents by employment status (unemployed, part-time, employed) and compute the average values of their basic needs over five weekdays. We visualize the five basic needs whose values exhibit the most noticeable fluctuations. As shown in Figure~\ref{fig:basic_needs}, these attributes fluctuate least for unemployed agents, moderately for part-time workers, and most dramatically for employed agents. All attributes, except \textit{Fullness}, follow a consistent pattern: a steady decline between 9:00 and 18:00, followed by recovery during non-working hours. This trend arises because employed agents are predominantly occupied with work during the day,  which restricts them from engaging in replenishing activities. \textit{Fullness}, however, rises at 8:00, 12:00, and 18:00, corresponding to mealtimes.

\subsection{Transportation Statistics}

We additionally evaluate transportation mode preferences across demographic categories, as summarized in Table~\ref{tab:transportation}. 
Overall, walking emerges as the dominant choice across all groups, reflecting its suitability for short-distance travel. 
PMV usage remains consistently low, which aligns with mobility patterns observed in our ground-truth dataset. 
Agents tend to rely on walking for nearby destinations and switch to public transit for longer routes, resulting in a natural bimodal split between these two modes. 
Environmental factors also contribute: PMV is rarely selected during rainy periods due to reduced safety and comfort. 
Across all categories, the experimental results closely track the ground-truth percentages, indicating that our agent-based mobility model successfully captures realistic travel preferences.

\begin{table}[t]
\caption{Time percentage~(\%) spent by agents using different transportation modes. ``Exp'' represents experimental results from our simulation, while ``GT'' refers to ground truth values from our proprietary dataset.}
\centering
\resizebox{1.0\linewidth}{!}{
\begin{tabular}{ccccccc}
\hline
\textbf{Category} & \multicolumn{2}{c}{\textbf{Walking}} & \multicolumn{2}{c}{\textbf{PMV}} & \multicolumn{2}{c}{\textbf{Bus}} \\
\cline{2-7}
& \textbf{Exp} & \textbf{GT} & \textbf{Exp} & \textbf{GT} & \textbf{Exp} & \textbf{GT} \\
\hline
100 & 89.96 & 88.78 & 0.00 & 0.57 & 10.04 & 10.65 \\
110 & 93.99 & 92.74 & 0.00 & 0.43 & 6.01 & 6.83 \\
120 & 94.39 & 93.19 & 0.00 & 0.29 & 5.61 & 6.52 \\
210 & 96.84 & 95.59 & 0.87 & 1.54 & 2.29 & 2.87 \\
211 & 95.44 & 93.92 & 0.00 & 0.53 & 4.56 & 5.55 \\
220 & 95.41 & 94.36 & 0.00 & 0.56 & 4.59 & 5.08 \\
300 & 95.43 & 94.21 & 0.87 & 1.41 & 3.70 & 4.38 \\
310 & 92.43 & 90.98 & 3.23 & 3.99 & 4.34 & 5.04 \\
320 & 94.55 & 93.45 & 0.00 & 0.38 & 5.45 & 6.17 \\
\hline
\end{tabular}}
\label{tab:transportation}
\end{table}

\section{Conclusion}
We presented \textbf{MobileCity}, a scalable framework for large-scale generative-agent simulation in dynamic urban environments. Our system integrates a realistic multi-modal transportation model and a unified agent architecture that jointly incorporates static human parameters, dynamic basic needs, temporal habits, and compulsory tasks. Through asynchronous batched action selection and lightweight communication based on memory exchange, MobileCity achieves human-like behavioral realism while remaining computationally efficient. The resulting simulations provide fine-grained insights into urban mobility patterns, offering a practical tool for improving traffic safety, infrastructure design, and urban planning while reducing reliance on costly real-world data collection.

\section{Limitations}
There are several limitations to our work. First, our simulation framework primarily focuses on modeling typical urban scenarios, while rare or extreme events, such as natural disasters, rapid population shifts, or sudden infrastructure failures, remain challenging to accurately capture. Second, the computational demands of large-scale, high-resolution urban simulations may become costly. Trade-offs in spatial granularity, temporal resolution, or agent complexity are necessary, which may limit the ability to represent micro-scale dynamics or long-term urban evolution. Besides, the behavior of agents may inherit biases present in the underlying data or model training. This includes reproducing social, cultural, or policy biases, as well as occasional generation of inconsistent or unfounded outputs. Finally, our work raises ethical and policy considerations. Automated urban simulations have the potential to influence real-world decision-making. It is therefore critical that users remain aware of the inherent biases and limitations of these systems.

\section{Ethics Statement}

This paper presents MobileCity, an LLM-powered agent framework designed to simulate large-scale urban mobility and social behaviors in a cost-effective and scalable manner. While our approach offers significant benefits for urban planning, traffic management, and behavioral modeling, it also raises several ethical considerations.  

One primary concern is the potential for bias amplification. Since our agent behaviors are derived from survey data and LLM-generated actions, any biases inherent in these sources could propagate within the simulation. This may lead to an unrealistic or skewed representation of population behaviors, which, if used for policy-making or infrastructure design, could reinforce existing social or economic inequalities.  

Another potential risk is the misuse of simulation insights. The ability to predict crowd density, individual behaviors, and mobility trends may be leveraged for unethical purposes, such as excessive surveillance, behavioral manipulation, or commercial exploitation without public consent. Safeguards should be in place to ensure that data-driven insights are used responsibly and in ways that benefit society.  

To mitigate these risks, we advocate for the responsible deployment of our framework, emphasizing transparency, fairness, and the inclusion of human oversight when deriving actionable insights from the simulation. By adhering to these principles, we can ensure that the use of generative agents in urban simulations remains ethically and socially beneficial.

\bibliography{custom}

\clearpage
\appendix

\begin{strip}
\centering
\includegraphics[width=\textwidth]{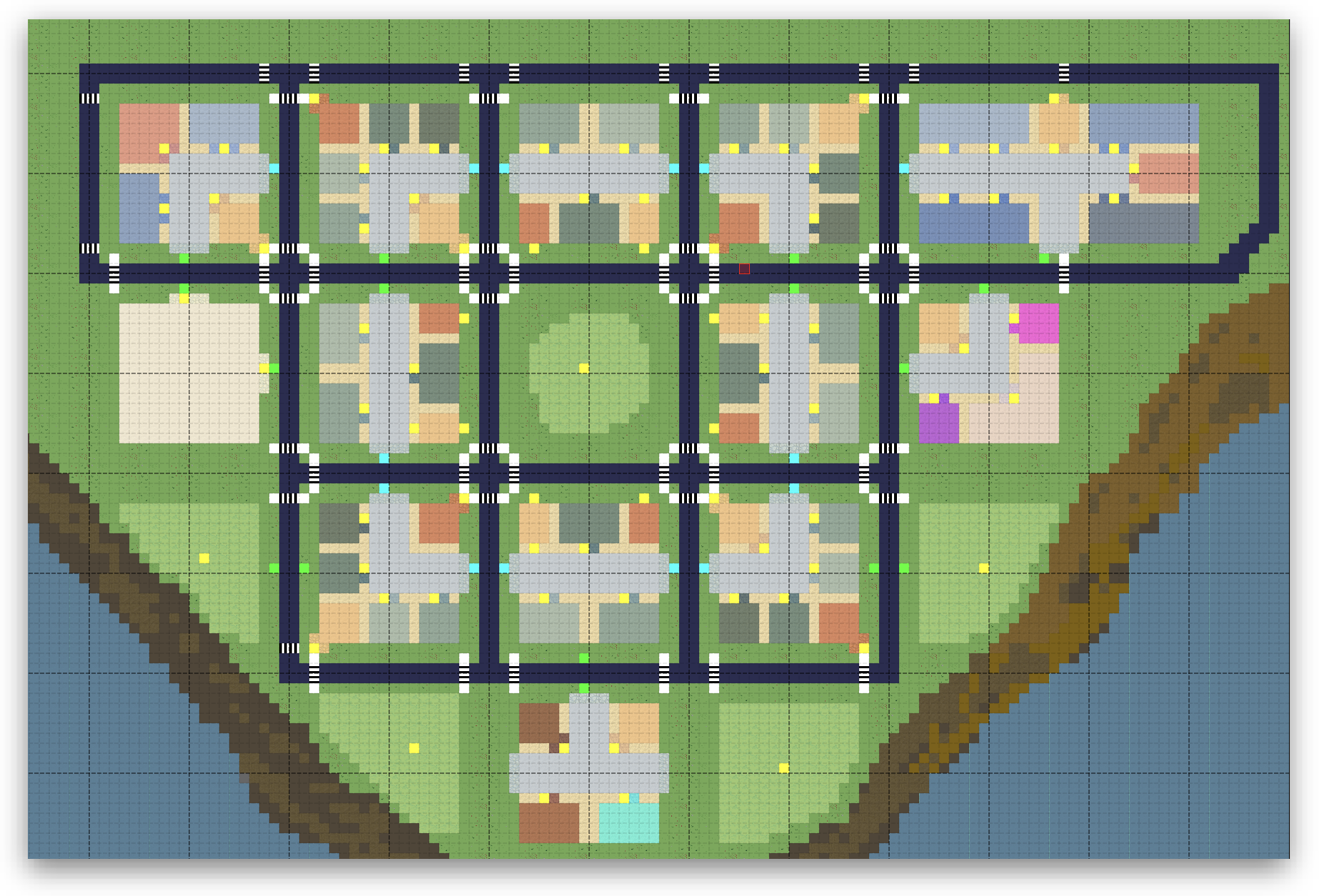}
\captionof{figure}{The map of our simulated city.}
\label{fig:map}
\end{strip}

\section{City Simulator}
In our simulation, agents operate in discrete time steps of 15 seconds, during which they perform actions, move between venues, and perceive their environment. To enhance scalability and efficiency, we introduce three key improvements over traditional simulators~\cite{uist/ParkOCMLB23, corr/abs-2402-02053}. First, to accommodate large-scale agent populations our simulator allows stateless simulation, significantly reducing computational overhead. Second, our platform includes diverse buildings and venues, enabling a more comprehensive representation of urban environments. Third, we design a transportation system with multiple mobility hubs, offering agents diverse route options and facilitating citywide movement patterns.

\subsection{City Map}
\label{sec:appendix_map}
\begin{figure}[h]
\vspace{-0.2cm}
    \begin{center} 
    \centering 
        \begin{subfigure}[b]{\linewidth} 
            \centering 
            \includegraphics[width=\linewidth]{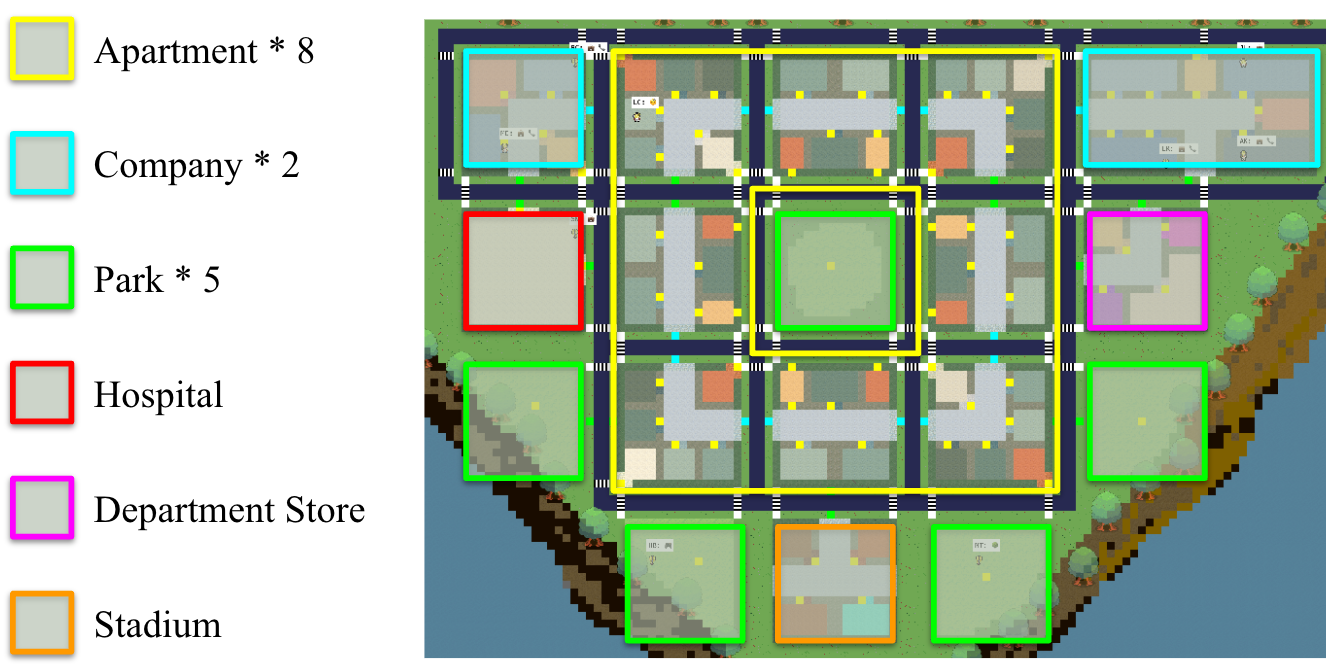} 
            \label{fig:buildings} 
        \end{subfigure} 
        \vspace{-0.5cm}
        \begin{subfigure}[b]{\linewidth} 
            \centering 
            \includegraphics[width=\linewidth]{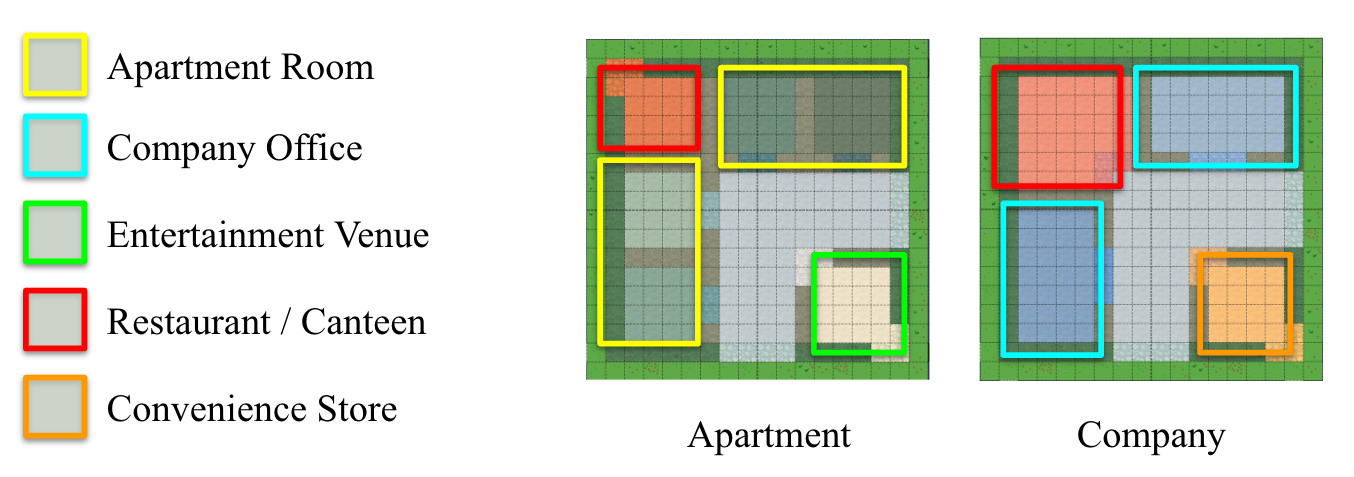} 
            \label{fig:rooms} 
        \end{subfigure} 
        \caption{The buildings and venues in the simulated city.} 
        \label{fig:buildings_rooms} 
    \end{center} 
\end{figure}

Our simulated city is constructed using a tile-based map representation, as shown in Figure~\ref{fig:map}. The city map follows a grid-based structure where each tile represents 25 meters, and each block spans 500 meters. It features a diverse range of urban infrastructures, including 18 buildings,  and 68 venues~(Figure~\ref{fig:buildings_rooms}). The city features 8 apartment complexes, 2 company offices, 5 parks, 1 hospital, 1 department store, and 1 stadium. Each building contains different spaces. For example, an apartment building contains several living spaces for agents, 1 restaurant, and 1 convenience store or 1 entertainment venue. A company building contains several offices, 1 company canteen, and 1 convenience store.

\begin{strip}
    \centering
    \includegraphics[width=14cm]{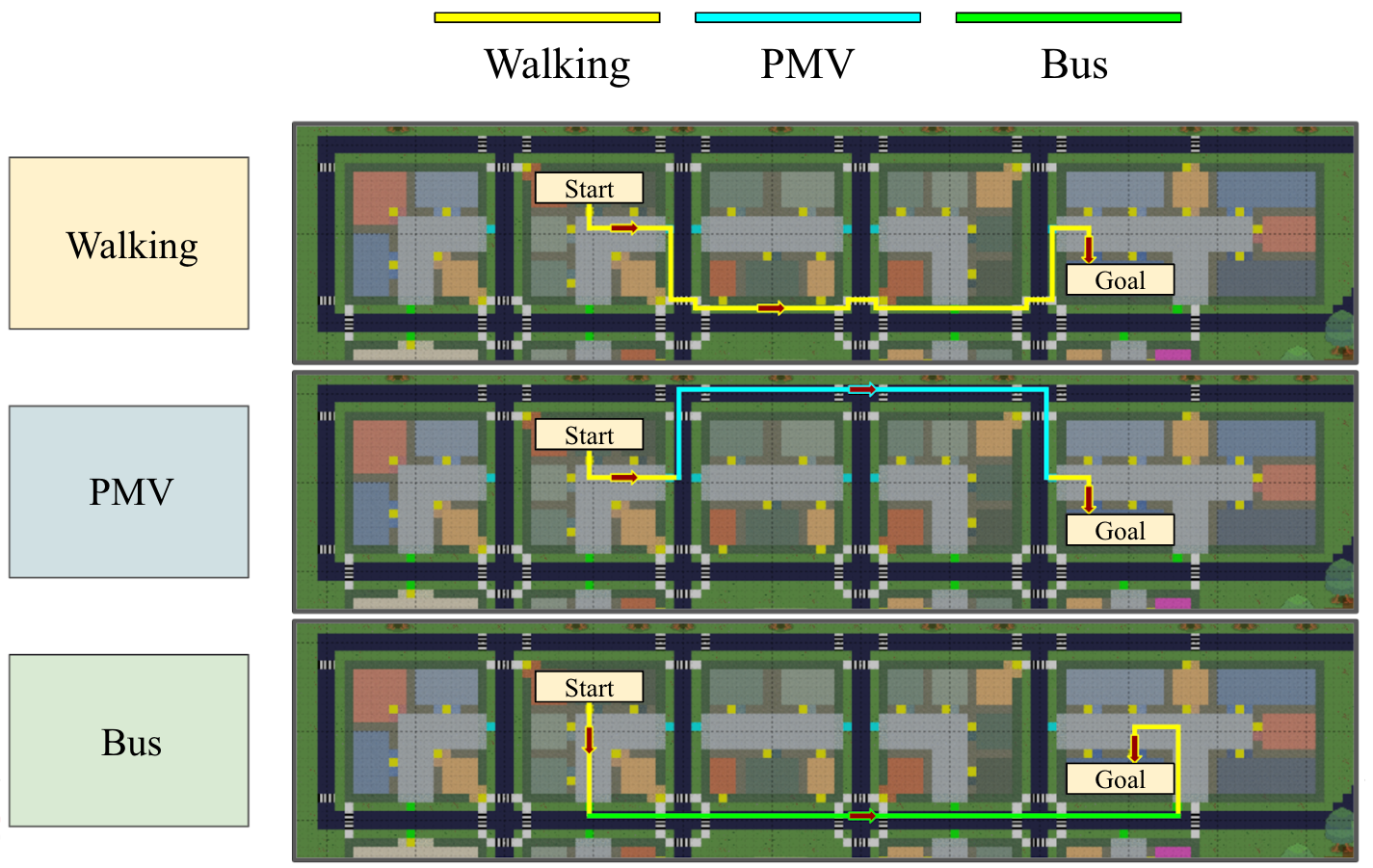}
    \captionof{figure}{An agent traveling from apartment to company has three route options: (1) walking through zebra crossings (yellow lines), (2) walking to PMV (\textbf{P}ersonal \textbf{M}obility \textbf{V}ehicle) station, riding on highway, then walking to destination (blue lines), or (3) walking to bus station, taking bus, then walking to destination (green lines).}
    \label{fig:routes}
\end{strip}

\subsection{Transportation System}
\label{sec:appendix_transportation}
The diversity of transportation modes facilitates the investigation of mobility patterns among urban residents~\cite{abs-2112-12071}. In our simulation, agents move using three transportation modes: Walking, PMV, and Bus, where PMV refers to a personal mobility vehicle. We introduce a constrained navigation system that dynamically determines optimal routes based on cost, constraints, and individual preferences. Inspired by real-world systems~\cite{wardman2020generalised}, our navigation system generates multiple route options, each differing in time cost and monetary cost. In general, bus routes have the lowest time cost but the highest monetary cost, whereas walking routes are the opposite. To formalize this, we construct three graphs \cite{corr/abs-1809-05481}: a walking graph $G_w$, a PMV graph $G_p$, and a bus graph $G_b$ in our map. Each graph is constructed with nodes representing accessible points for agents, and edges representing different moving costs. 

\begin{figure}[h]
\begin{center}
\centerline{\includegraphics[width=\linewidth]{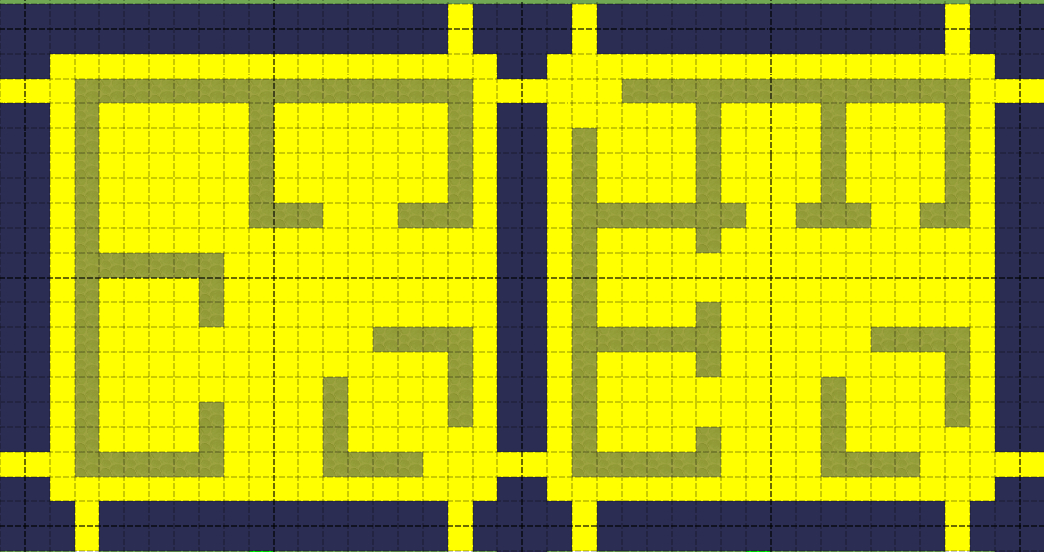}}
\caption{Walking-accessible zone.}
\label{fig:walking_graph}
\end{center}
\vspace{-1cm}
\end{figure}

\begin{figure}[h]
\begin{center}
\centerline{\includegraphics[width=\linewidth]{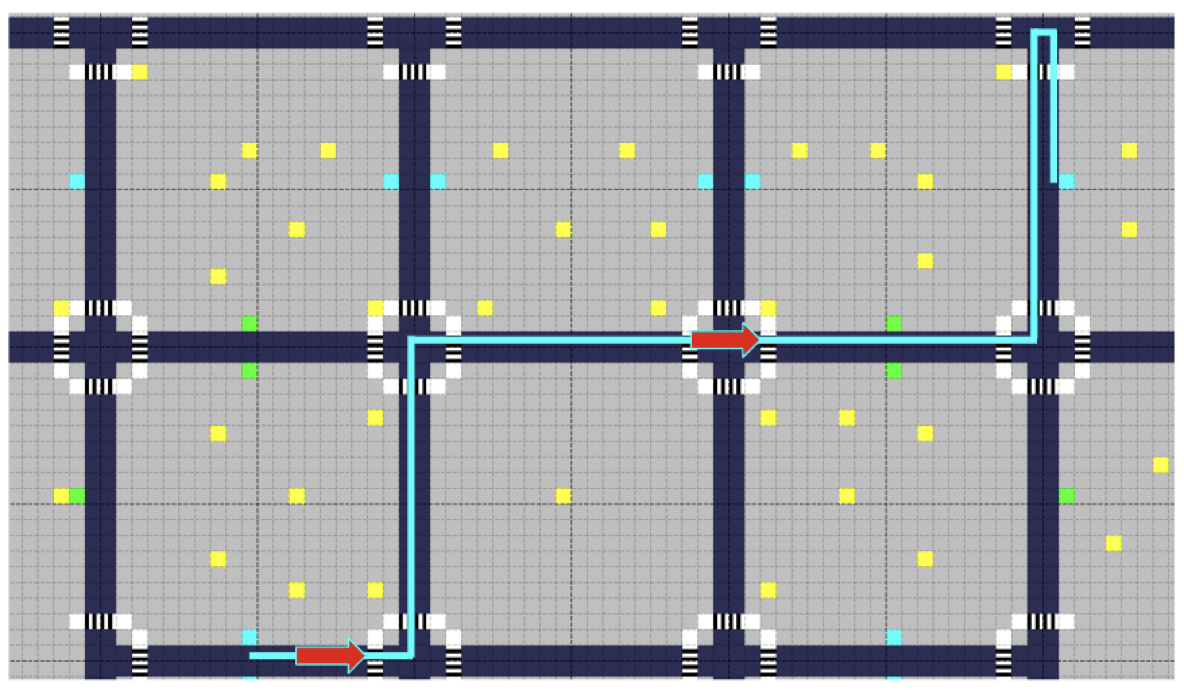}}
\caption{A PMV route example.}
\label{fig:PMV_graph}
\end{center}
\vspace{-1cm}
\end{figure}

A walking graph consists of passages inside buildings, which are yellow areas in Figure~\ref{fig:walking_graph}, and zebra crossings between buildings in Figure~\ref{fig:PMV_graph}. In one building, agents can access most of the areas except for collision walls. Between buildings, agents can walk across zebra crossings on highways. An agent moves 1 tile in each time step by walking.

A PMV graph consists of nodes of PMV stations, represented as blue tiles in Figure~\ref{fig:PMV_graph}. To ride a PMV, agents must walk to the PMV station first, then ride the PMV on the left side of the highway. An agent moves 2 tiles in each time step when using a PMV.

\begin{figure}[h]
    \centering
    \includegraphics[width=\linewidth]{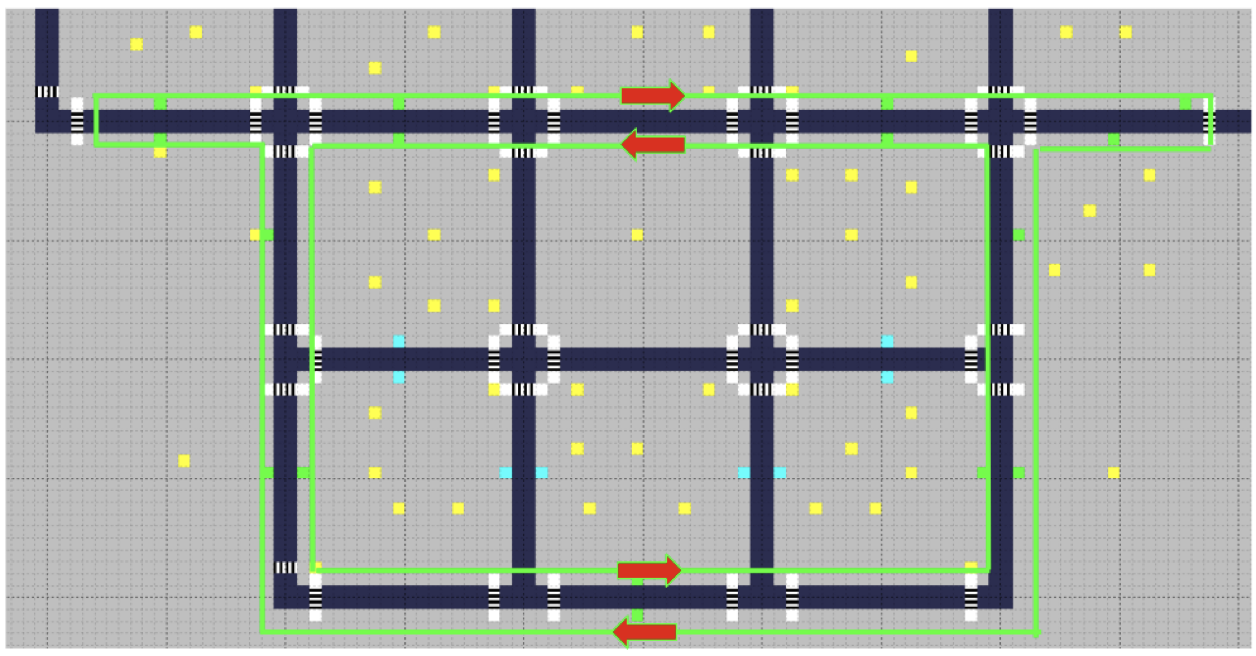}
    \caption{Two bus routes in our city.}
    \label{fig:bus_graph}
\end{figure}

A bus graph consists of nodes of bus stations, represented as green tiles in Figure~\ref{fig:bus_graph}. To get on a bus, agents must walk to the bus station first, and then the bus will move on the left side of the highway. An agent moves 5 tiles in each time step when using a bus.

At each simulation time step, agents traverse 1, 2, or 5 tiles depending on whether they are walking, using a PMV, or taking a bus, respectively. Consequently, the time required to travel across a full block is 300, 150, and 60 seconds for walking, PMV, and bus travel, respectively.

Their respective time costs $t_w$, $t_p$, and $t_b$ are calculated as:
\begin{align*}
& t_w = \min_{\pi \in G_w} \text{dist}(s \rightarrow t), \\
& t_p = \min_{\pi \in G_w \cup G_p} \text{dist}(s \rightarrow t), \\
& t_b = \min_{\pi \in G_w \cup G_b} \text{dist}(s \rightarrow t),
\end{align*}
where $s$ is the starting place, $t$ is the terminal place, and $\pi$ represents all the paths in graphs. Route selection is constrained by its upcoming compulsory tasks and influenced by agent group characteristics. For instance, if an agent must reach the office within 15 minutes, it prioritizes the bus to minimize travel time and avoid tardiness. Higher-income agents are more likely to choose bus due to its time efficiency, whereas lower-income agents may opt for walking to reduce costs. Additionally, weather conditions~\cite{weather} play a crucial role in mobility decisions, on rainy days, agents tend to avoid PMVs due to safety and comfort concerns.

\section{Individual Action Module}
\label{appendix:action}
We now explain the details in the action module.

\begin{figure*}[t]
    \centering
    \includegraphics[width=15cm]{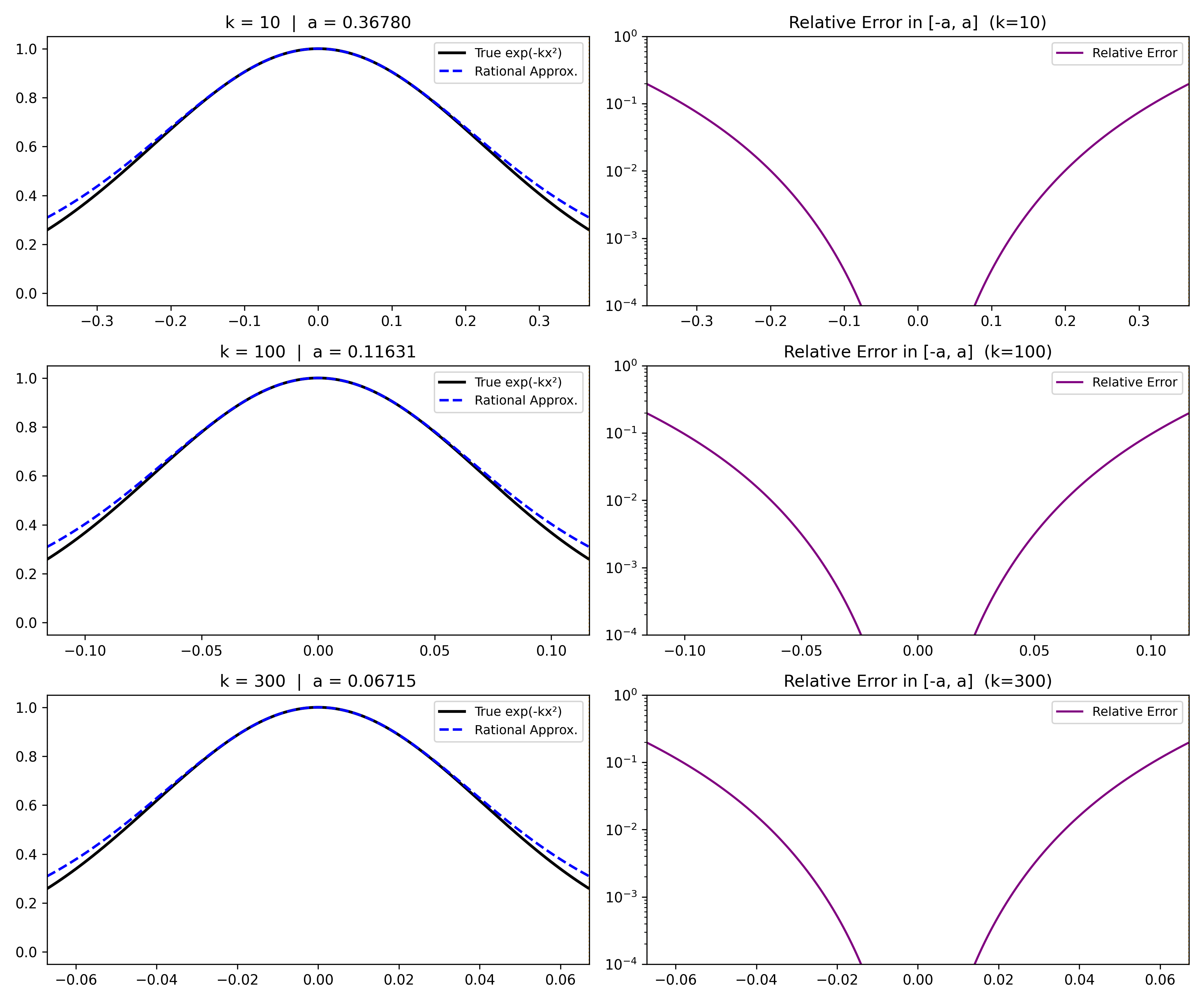}
    \caption{The curve remains close to zero throughout the interval, demonstrating that the Padé [2/2] approximation tracks the exponential decay very accurately. The maximum relative error is among 0.1\% and 1\%.}
    \label{fig:gaussian_rational}
    \vspace{-0.5cm}
\end{figure*}

\subsection{Needs-driven Action}
\label{sec:appendix_needs}
The total needs-based score at time \(t\) is defined as:
\begin{equation}
N(t) = N_{\mathrm{hp}} \, N_{\mathrm{imp}}
\label{eq:needs_total}
\end{equation}

Let \(x_{\mathrm{hp}} \in \mathbb{R}^D\) denote the agent’s human parameter (HP) vector, 
and \(x_{\mathrm{act}} \in \mathbb{R}^D\) the action’s HP vector, with weights \(w \in \mathbb{R}_{\ge 0}^D\).
The weighted cosine similarity is given by:
\begin{equation}
N_{\mathrm{hp}}
= \frac{1}{2}\Bigg(1 + 
\frac{\langle w\!\odot\!x_{\mathrm{hp}},\, w\!\odot\!x_{\mathrm{act}}\rangle}
{\|w\!\odot\!x_{\mathrm{hp}}\|_2\,\|w\!\odot\!x_{\mathrm{act}}\|_2}
\Bigg),
\label{eq:N_hp}
\end{equation}
which maps the similarity to the range \([0,1]\).

Let:
\begin{itemize}
  \item \(I_N \in \mathbb{R}^8\): the agent’s importance weights for each of the 8 needs;
  \item \(C_N \in [0,1]^8\): the agent’s current need satisfaction levels (scaled to \([0,1]\));
  \item \(A_N \in \mathbb{R}^8\): the action’s positive contribution to each need.
\end{itemize}

Then the importance-based need score is defined element-wise as:
\begin{align}
N_{\mathrm{imp}} 
&= 
\big\langle
\operatorname{softmax}(I_N)
\odot (1 - C_N)
\nonumber\\[-2pt]
&\quad\odot\;
\tanh\!\big(k_{\tanh}\,\operatorname{ReLU}(A_N)\big)
\big\rangle
\label{eq:N_imp}
\end{align}

where:
\begin{itemize}
  \item \(\operatorname{softmax}(I_N)\) normalizes the importance of each need;
  \item \((1 - C_N)\) represents the current deficiency or gap in satisfaction;
  \item \(\tanh(k_{\tanh}\,\operatorname{ReLU}(A_N))\) introduces diminishing returns on positive need fulfillment,
        ensuring saturation as contribution increases.
\end{itemize}

\subsection{Habit-driven Action}
\label{sec:appendix_habit}
The habit-based score at time $t$ is defined as:
\begin{equation}
H(t) = R(t)A_H \, \exp\!\big(-\,k_H\,\Delta\theta(t)^{2}\big)
\label{eq:habit_gaussian}
\end{equation}

$R(t)$ will be neglected in the following discussion since it's not related to Gaussian distribution.

Our rationale for modeling habit strength using a Gaussian distribution is as follows. First, habit strength is treated as a continuous variable rather than a binary one. Habit strength is the cumulative result of numerous minor factors in reality. In behavioral prediction and health psychology, research~\cite{rebar2019we} has found that variables such as intention, behavior, and frequency are ``approximately normally'' distributed. Therefore, according to the Central Limit Theorem, the aggregation of these influences will approximate a normal distribution.

Given this premise, we discuss the derivation of the formula for $A_H$ and $k_H$. 

\textbf{Object 1}: given a half-action duration $a$, the integral of the habit strength over the interval $[-a, a]$ must account for more than 90\% of the total integral area.

Consider a normalized Gaussian kernel:
\begin{equation}
f(x) = e^{-k_H x^2}, \quad k_H > 0.
\end{equation}
The total area under this curve is:
\begin{equation}
\int_{-\infty}^{\infty} e^{-k_H x^2}\,dx = \sqrt{\frac{\pi}{k_H}}.
\end{equation}
When integrating over a finite range $[-a, a]$, the result becomes:
\begin{equation}
I(a) = \int_{-a}^{a} e^{-k_H x^2}\,dx = \sqrt{\frac{\pi}{k_H}}\,\mathrm{erf}(\sqrt{k_H}a).
\label{eq:gauss_integral}
\end{equation}

The error function is defined as:
\begin{equation}
\mathrm{erf}(x) = \frac{2}{\sqrt{\pi}}\int_0^x e^{-t^2}\,dt.
\end{equation}
By substituting $t = \sqrt{k_H}x$, the Gaussian integral over $[-a,a]$ introduces $\mathrm{erf}(\sqrt{k_H}a)$.  
The fraction of total area within $[-a,a]$ is therefore:
\begin{equation}
P(a) = 
\frac{\int_{-a}^{a} e^{-k_H x^2}\,dx}{\int_{-\infty}^{\infty} e^{-k_H x^2}\,dx}
= \mathrm{erf}(\sqrt{k_H}a).
\label{eq:ratio}
\end{equation}

In our experimental setup, the execution duration of non-mandatory tasks ranges from 0.5 to 3 hours, and correspondingly $a \in [\pi/48, \pi/8]$. 
If we require that $[-a,a]$ contains 90\% of the total area, we solve:
\begin{equation}
\mathrm{erf}(\sqrt{k_H}a) = 0.9.
\end{equation}
This yields:
\begin{equation}
\sqrt{k_H}a = \mathrm{erf}^{-1}(0.9) \approx 1.163,
\end{equation}
and consequently:
\begin{equation}
k_H \approx \left(\frac{1.163}{a}\right)^2.
\label{eq:k_a_relation}
\end{equation}

Since the computation of the exponential function is computationally expensive, we perform a rational approximation.

\textbf{Object 2}: given $k_H$, the habit peak $A_H$ must vary to ensure the integral remains constant.

\begin{equation}
I(k_H) = A_H \int_{-a\sqrt{k_H}}^{a\sqrt{k_H}} e^{-t^2}\,\frac{dt}{\sqrt{k_H}}
     = 0.9*A_H\,\frac{\sqrt{\pi}}{\sqrt{k_H}}
\end{equation}

To maintain a constant area \(S\):
\begin{equation}
A_H(k_H) \approx 0.627\,S\,\sqrt{k_H}
\end{equation}

Since the computation of the exponential function is computationally expensive, we apply Padé approximation that preserves accuracy near \(u = 0\).

\begin{equation}
e^{-u} \approx \frac{1 - \tfrac{u}{2} + \tfrac{u^2}{12}}{1 + \tfrac{u}{2} + \tfrac{u^2}{12}}.
\end{equation}

By substituting \(u = k_Hx^2\), we obtain the rational approximation of \(f(x)\):
\begin{equation}
f(x) \approx
A_H\,\frac{1 - \tfrac{k_Hx^2}{2} + \tfrac{k_H^2x^4}{12}}
{1 + \tfrac{k_Hx^2}{2} + \tfrac{k_H^2x^4}{12}}.
\end{equation}

The maximum relative error is among 0.1\% and 1\%, as demonstrated in Fig.~\ref{fig:gaussian_rational}.

\subsection{Obligatory-driven Action}
\label{sec:appendix_obilgatory}

\(\mathrm{Mask}(t)\) is True if and only if three conditions are satisfied, 

\begin{equation}
\mathrm{Mask}(t) = M_{\text{sem}}(t,a_{\text{act}})\;M_{\text{open}}(t,a_{\text{act}})\;M_{\text{obl}}(t,a_{\text{act}})
\end{equation}

\textbf{Semantic–temporal consistency} ensures that the action’s semantics align with the current time period. An action labeled \textit{eat breakfast} should be invalid in the evening. $M_{\text{sem}}(t,a_{\text{act}})=1$ requires

\begin{equation}
t \in T_{\text{act}}^{\text{sem}}
\end{equation}

\textbf{Venue availability} ensures that the physical location associated with an action must be open during the planned execution interval. Let  $\Delta t_{\text{cur}}(a_{\text{act}})$ the travel time from the current location to the next action location. $M_{\text{open}}(t,a_{\text{act}})=1$ requires

\begin{align}
t + \Delta t_{\text{cur}}(a_{\text{act}}) &\ge t_{\text{start}}^{\text{venue}}, \label{eq:open1}\\[3pt]
t + \Delta t_{\text{cur}}(a_{\text{act}}) + \Delta t_{\text{act}} &\le t_{\text{close}}^{\text{venue}}. \label{eq:open2}
\end{align}

\textbf{Obligation constraint} ensures that the agent must complete all ongoing voluntary actions before the next scheduled mandatory task.  Let $t_{\text{next}}^{\text{obl}}$ denote the start time of the next obligation, and $\Delta t_{\text{next}}(a_{\text{act}})$ the travel time from the current location to the next mandatory task location. $M_{\text{obl}}(t,a_{\text{act}})=1$ requires

\begin{equation}
        t + \Delta t_{\text{cur}}(a_{\text{act}})
        + \Delta t_{\text{act}}
        + \Delta t_{\text{next}}(a_{\text{act}})
        \le t_{\text{next}}^{\text{obl}}
\end{equation}

\subsection{Action Selector}
\label{appendix_action_selector}
We now provide a comprehensive explanation of Agent Action Selector, detailing the implementation and technical details. This is a detailed example.

\begin{tcolorbox}[colframe=blue!40!black, colback=white, title=Action Selector,breakable]
Now, it is 7:00 AM on Monday, and our agent Alex Kim wakes up. He will select an action by following these steps.\\

\textbf{Step 1:} Consider restraints from the next Obligatory-driven Action.
Alex is a 25-year-old software engineer working. He needs to start working remotely or in the office from 9:00 on weekdays.\\

\textbf{Step 2:} List Needs-driven Actions. Alex needs to eat a lot to maintain energy for high-intensity work, which means his has a high demand for needs of ``Fullness'' and ``Energy''. He is very hungry, so his Top-5 needs-driven actions will be: \textit{have breakfast in the canteen}, \textit{grasp some food from the convenience store}, \textit{drink coffee in the cafe}, \textit{have decent breakfast at a nearby restaurant}, and \textit{have breakfast at home}.\\

\textbf{Step 3:} List Habit-driven Actions. According to his personal habits, Alex's Top-3 actions at 7:00 are:  \textit{drink coffee in the cafe}, \textit{walk in the park}, and \textit{meditate at home}.\\

\textbf{Step 4:} Select an action and transportation mode with LLM. The current environmental condition is: sunny, 15°C. It's a good weather to go out, LLM makes the action choice for Alex: \textit{drink coffee in the cafe}. Meanwhile, it takes 20 minutes to walk to the cafe, but only 8 minutes by PMV. LLM makes the transporation choice for Alex: \textit{PMV}.\\
\end{tcolorbox}

\section{Temporal Optimization}

\subsection{Asynchronous Actions}
\label{appendix:asyn_actions}

At every iteration, the simulator scans through the active agents. For each agent $a_i$, the system first checks whether the next event on its schedule 
is a mandatory task. If so, the agent executes that task immediately, updates its local time 
$\theta_i \leftarrow \theta_i + \Delta t_{\text{act}}$, 
and adjusts its need satisfaction vector 
$C_N \leftarrow \mathrm{clip}(C_N + A_N,\,[0,1])$,
where $A_N$ denotes the need-specific increments contributed by the action.
Otherwise, the agent’s action selector compiles two sets of candidate actions, 
$\textsc{Act}_{\text{needs}}$ from the needs-driven module 
and $\textsc{Act}_{\text{habit}}$ from the habit-driven module, 
and merges them into a unified candidate set \textsc{Cands}. Each candidate set, together with the agent’s persona, current environment view, 
and current need satisfaction vector $C_N$, is assembled into an LLM request. Instead of invoking the model immediately, the task is placed into a shared 
\textit{batch buffer}. When the batch size reaches a predefined threshold $B$, all queued tasks are sent to the LLM simultaneously as a parallel API call. The results are then returned asynchronously, and each agent updates its state independently according to the selected action. After each execution, if the local time $\theta_i$ of an agent reaches 24:00, the agent is temporarily removed from the active list.

\begin{algorithm}[h]
\caption{Asynchronous Batched Action Scheduling}
\label{alg:async-short}
\begin{algorithmic}[1]
\State Initialize agents $\mathcal{A}=\{a_1,\dots,a_N\}$ and clocks $\mathcal{I}=\{\theta_1,\dots,\theta_N\}$; $\mathcal{B}\gets\emptyset$
\While{$\mathcal{A}\neq\emptyset$}
  \For{each $a_i\in\mathcal{A}$}
    \If{$\theta_i\ge24{:}00$} remove $a_i$;\textbf{continue}
    \ElsIf{mandatory$(a_i)$} execute; $\theta_i\!+=\!\Delta t_{\text{act}}$; $C_N\!+=\!A_N$;\textbf{continue}
    \Else
      $\textsc{Cands}\!\leftarrow\!\textsc{MergeTopK}(\textsc{Act}_{\text{needs}},\textsc{Act}_{\text{habit}})$;\\
      add $(a_i,\textsc{Cands},C_N,\text{persona},\text{env})$ to $\mathcal{B}$; mark $a_i$ as awaited
    \EndIf
  \EndFor
  \If{$|\mathcal{B}|\!\ge\!B$ \text{ or all agents awaited}}
     dispatch $\mathcal{B}$ to LLM in parallel;\\
     update $\theta_i\!\leftarrow\!\theta_i+\Delta t_{\text{act}}$, $C_N\!\leftarrow\!\mathrm{clip}(C_N+A_N,[0,1])$; reset awaited flags; $\mathcal{B}\!\leftarrow\!\emptyset$
  \EndIf
\EndWhile
\end{algorithmic}
\end{algorithm}

\subsection{Asynchronous Conversations}
\label{appendix:asyn_conversations}
At every iteration, the simulator scans through the active agents and identifies potential conversation pairs $(a_i, a_j)$. 
Two types of conversations are considered:
(i)~\textbf{Face-to-face interactions} occur when two agents occupy the same venue within overlapping time windows, and 
(ii)~\textbf{Socially initiated communications} occur when an agent’s social need in $C_N$ exceeds a threshold and it proactively contacts another agent through a virtual channel. 

Each conversation pair is converted into a communication task 
$\textsc{Task}_{\text{conv}} = (a_i, a_j, \textsc{Memory}_i, \textsc{Memory}_j, \text{context}_{ij})$, 
where $\textsc{Memory}_i$ and $\textsc{Memory}_j$ denote the recent memory slots of each participant. 
Rather than invoking the LLM for each pair independently, the simulator appends these tasks to a global batch $\mathcal{B}_{\text{conv}}$. 
When the batch size reaches the threshold $B_{\text{conv}}$, all tasks are dispatched in parallel as a single batched API call: 
$\textsc{DispatchBatch}(\mathcal{B}_{\text{conv}}) = 
\big\{(\Delta \mathcal{M}_i, \Delta \mathcal{M}_j, \Delta R_{ij}) = 
\textsc{LLM}_{\text{comm}}(\textsc{Task}_{\text{conv}})\big\}$.
Here, $\Delta \mathcal{M}_i$ and $\Delta \mathcal{M}_j$ represent the exchanged memory indices, 
and $\Delta R_{ij}$ updates the bilateral relationship score between agents $i$ and $j$. 
Once processed, the updated memories and relationship states are written back into each agent’s local store: 
$\mathcal{M}_i \leftarrow \mathcal{M}_i \cup \Delta \mathcal{M}_i$, 
$\mathcal{M}_j \leftarrow \mathcal{M}_j \cup \Delta \mathcal{M}_j$, and 
$R_{ij} \leftarrow R_{ij} + \Delta R_{ij}$.

\begin{algorithm}[h]
\caption{Asynchronous Batched Conversation Scheduling}
\label{alg:async-conv}
\begin{algorithmic}[1]
\State Initialize active agents $\mathcal{A}$; conversation batch $\mathcal{B}_{\text{conv}}\!\gets\!\emptyset$
\While{$\mathcal{A}\neq\emptyset$}
  \For{each potential pair $(a_i,a_j)$ from $\mathcal{A}$}
    \If{face\_to\_face$(a_i,a_j)$ or high\_social\_need$(a_i)$}
        add $(a_i,a_j,\textsc{Memory}_i,\textsc{Memory}_j,\text{context}_{ij})$ to $\mathcal{B}_{\text{conv}}$
    \EndIf
  \EndFor
  \If{$|\mathcal{B}_{\text{conv}}|\!\ge\!B_{\text{conv}}$}
     dispatch $\mathcal{B}_{\text{conv}}$ to LLM in parallel;\\
     update $\mathcal{M}_i\!\leftarrow\!\mathcal{M}_i\cup\Delta\mathcal{M}_i$, 
     $\mathcal{M}_j\!\leftarrow\!\mathcal{M}_j\cup\Delta\mathcal{M}_j$, 
     $R_{ij}\!\leftarrow\!R_{ij}+\Delta R_{ij}$; 
     $\mathcal{B}_{\text{conv}}\!\leftarrow\!\emptyset$
  \EndIf
\EndWhile
\end{algorithmic}
\end{algorithm}

\section{Experiments}
\subsection{Dataset Description}
Our proprietary dataset is derived from a survey of over 4,000 respondents and contains continuous human parameters, ranging between 0 and 1. Human parameters capture personality and lifestyle traits. In addition, our dataset includes detailed daily activity schedules for each individual, specifying the modes of transportation used for different activities. These real-world schedules serve as a benchmark to assess the faithfulness of our proposed simulation framework, ensuring that it accurately reflects human behavioral patterns.

\subsection{Example of Daily Plans}
\label{sec:action_example}
An example of actions generated by baseline~\cite{uist/ParkOCMLB23} is provided below:
\begin{quote}
\small
\begin{minipage}{\linewidth}
\begin{verbatim}
00:00 sleeping
06:00 waking up, getting ready for the day 
06:30 having breakfast, checking her emails 
07:00 commuting to Hobbs Cafe
\end{verbatim}
\end{minipage}
\end{quote}
which receives a score of 3. And the actions generated by our model are:
\begin{quote}
\small
\begin{minipage}{1.0\linewidth}
\begin{verbatim}
07:06 wake up, stretch, make coffee
08:00 check messages, read the news
09:15 work on a project, attend virtual meeting
11:24 cook lunch, eat with a friend, 
chat with Mike
\end{verbatim}
\end{minipage}
\end{quote}
which scores 4 out of 5. 

To analyze agents’ emotional states, we group agents by employment status (unemployed, part-time, employed) and compute the average values of their basic needs over five weekdays. As shown in Figure~\ref{fig:basic_needs}, the attributes fluctuate least for unemployed agents, moderately for part-time workers, and most dramatically for employed agents. All attributes, except \textit{fullness}, follow a consistent pattern: a steady decline between 9:00 and 18:00, followed by recovery during non-working hours. This trend arises because employed agents are predominantly occupied with work during the day,  which restricts them from engaging in replenishing activities. \textit{Fullness}, however, rises at 8:00, 12:00, and 18:00, corresponding to mealtimes.

\subsection{Additional Baseline Information}
In this section, we present a comparative analysis of our proposed framework, \textit{MobileCity}, against three widely recognized approaches for modeling urban interactions: \textit{SmallCity} \cite{uist/ParkOCMLB23}, \textit{AGA} \cite{corr/abs-2402-02053}, and \textit{HumanoidAgent} \cite{emnlp/WangCC23}. Our evaluation focuses on six key dimensions essential for simulating real-world urban behaviors: \textit{daily activities}, \textit{long-term habits}, \textit{basic needs}, \textit{remote communication}, \textit{vehicle usage}, and \textit{movements}. Table \ref{tab:dataset_comparison} provides a detailed comparison of these methods with human behavior.

The \textit{Daily Activities} column assesses a system’s capacity to execute structured, day-to-day tasks, while \textit{Long-Term Habits} measures its ability to develop and sustain recurring behavioral patterns over time. The \textit{Basic Needs} criterion reflects the model’s capability to account for essential human necessities. \textit{Remote Communication} evaluates how well the system facilitates interactions with external entities across distances. \textit{Vehicle Usage} examines mobility-related functionalities, and \textit{Compulsory Tasks} refers to the model’s ability to incorporate mandatory or routine obligations into its behavioral framework.
\begin{table*}[t]
\centering
\caption{Comparison of \textit{MobileCity} with prior approaches. }

\begin{tabular}{lcccccc}
\toprule
\multirow{2}{*}{Name} & Daily & Long-Term & Basic & Remote & \multirow{2}{*}{Vehicles} & Compulsory \\
 & Activities & Traits & Needs & Communication &  & Tasks \\
\midrule
SmallCity & \ding{51} & \ding{55} & \ding{55} & \ding{55} & \ding{55} & \ding{55} \\
AGA & \ding{51} & \ding{55} & \ding{55} & \ding{55} & \ding{55} & \ding{55} \\
HumanoidAgent & \ding{51} & \ding{55} & \ding{51} & \ding{55} & \ding{55} & \ding{55} \\
MobileCity & \ding{51} & \ding{51} & \ding{51} & \ding{51} & \ding{51} & \ding{51} \\
\bottomrule
\end{tabular}
\label{tab:dataset_comparison}
\end{table*}

\section{Discussion}
\subsection{Potential Improvement}
Our model presents several directions for future enhancement.

First, the introduction of rare events represents a significant challenge. While we have enhanced the plausibility of agent behaviors through the implementation of both dynamic and static agent characteristics, our current framework does not account for environmental dynamic     n variations beyond weather patterns. To investigate collective behavioral patterns during emergency scenarios such as earthquakes, floods, or fires, these events would need additional modules to produce human-like responses.

Second, our agent interaction mechanisms require refinement. The current paradigm restricts interactions to conversations between agents. A more valid approach would permit multi-agent dialogue sessions and collective activities such as group recreational events.

Third, the model does not yet fully represent heterogeneity in behavioral execution. In real settings, agents require varying durations to complete the same actions, and the resulting attribute changes differ across individuals. Future work should more precisely formalize and parameterize the relationship between agent personality traits and the variability in behavioral outcomes.
\subsection{Future Research}
Future research endeavors could concentrate on the following directions.

First, cross-cultural urban simulation represents a promising avenue of inquiry. The incorporation of cultural factors and their influence on urban agent behaviors would enable the exploration of divergent collective behavioral patterns across different cultural contexts. Additionally, the datasets serving as foundational sources should encompass subjects from diverse cultural backgrounds to ensure comprehensive representation.

Second, policy evaluation applications offer significant practical value. Leveraging simulation outcomes to assess the potential implications of urban planning decisions and to forecast behavioral adaptations among citizens following the implementation of various policies could inform evidence-based governance strategies.

Third, long-term memory and learning mechanisms require careful examination. Changes in the environment affect how agents accumulate and transfer experiences, shaping their future behaviors based on past interactions. For example, if a transportation route becomes congested due to infrastructure changes, and agents share this information within the system, a shift in commuting patterns is expected as agents adapt to avoid delays.

\clearpage

\end{document}